\documentclass[aps,prb,twocolumn,groupedaddress,showpacs,floatfix]{revtex4-1}

\usepackage{graphicx}
\usepackage{grffile}
\usepackage{subfigure}
\usepackage{amsmath}
\usepackage{amssymb}
\usepackage{bm}
\usepackage{nicefrac}

\graphicspath{{./}{figures/}}

\def\ket#1{\ensuremath{\vert{#1}\rangle}}

\begin{document}

\title{Fermi surface topology of LaFePO and LiFeP}

\author{Johannes Ferber}
\email{ferber@itp.uni-frankfurt.de}
\affiliation{Institut f\"ur Theoretische Physik, Goethe-Universit\"at Frankfurt, Max-von-Laue-Strasse 1, 60438 Frankfurt/Main, Germany}

\author{Harald O. Jeschke}
\affiliation{Institut f\"ur Theoretische Physik, Goethe-Universit\"at Frankfurt, Max-von-Laue-Strasse 1, 60438 Frankfurt/Main, Germany}

\author{Roser Valent\'\i}
\affiliation{Institut f\"ur Theoretische Physik, Goethe-Universit\"at Frankfurt, Max-von-Laue-Strasse 1, 60438 Frankfurt/Main, Germany}

\date{\today}

\begin{abstract}
  We perform charge self-consistent LDA+DMFT (density functional theory
  combined with dynamical mean field theory) calculations to study
  correlation effects on the Fermi surfaces of the iron pnictide
  superconductors LaFePO and LiFeP. We find a distinctive change in the topology
  of the Fermi surface in both compounds where a hole pocket with Fe~$d_{z^2}$ orbital character
  changes its geometry from a closed shape in LDA to an open shape upon inclusion of
  correlations. The opening of the pocket occurs in the vicinity of the $\Gamma$ (Z) point in LaFePO (LiFeP).
  We discuss the relevance of these findings for the low superconducting
  transition temperature and the nodal gap observed in these materials.

\end{abstract}


\maketitle


The iron pnictides are an intensively studied new class of
superconductors with superconducting transition temperatures up to
currently 56~K.~\cite{Wu2009} They show a variety of distinct properties 
in the normal as well as superconducting state like absence/presence
of magnetic ordering, weak to strong electronic correlations and
nodal/nodeless superconducting gaps. Most parent compounds exhibit a non-superconducting
ground state with antiferromagnetic order and become superconducting upon doping or
application of external pressure.

LaFePO is the iron pnictide in which superconductivity was
reported~\cite{Kamihara2006} for the first time at a critical temperature of about 6~K.
In LaFePO superconductivity arises without
doping or application of pressure, there is no long-range magnetic
order,~\cite{McQueen2008} and the superconducting gap is
nodal.~\cite{Yamashita2009, Fletcher2009, Hicks2009} As a
measure of electronic correlations, mass enhancement
values have been reported from a number of different experiments
including angle-resolved photoemission spectroscopy
(ARPES),~\cite{Lu2008} optical conductivity,~\cite{Qazilbash2009}
de~Haas-van~Alphen (dHvA),~\cite{Coldea2008} and specific heat
measurements;~\cite{Suzuki2009} all point to a range $m^\ast/m_{\rm
  LDA}\approx1.7\!-\!2.2$, consistent also with existing LDA+DMFT
studies.~\cite{Skornyakov2010, Yin2011}

Recently, LiFeAs received particular attention due to a number of
features that make it very attractive for both theoretical and
experimental studies: like LaFePO, it exhibits superconductivity
without doping or pressure at $T_c=18$~K~\cite{Tapp2008} and is nonmagnetic;
furthermore, it cleaves between adjacent Li layers mitigating the
issue of polar surfaces for surface-sensitive probes.

LiFeP shares these properties, yet its lower superconducting
transition temperature of 5~K~\cite{Deng2009} didn't trigger as much
research. LiFeP shows contrasting behavior to LiFeAs in some respects,
though. Most importantly, its superconducting order parameter is
nodal~\cite{Hashimoto2012} compared to the nodeless gap in
LiFeAs. This is unexpected as previous works suggested a relation
between the lack of a third hole pocket at the $\Gamma$
point and the formation of gap
nodes,~\cite{Kuroki2009,Ikeda2010,Thomale2011,Hirschfeld2011} but both LiFeAs and
LiFeP show three hole sheets at $\Gamma$. Effective masses in LiFeP were
extracted from resistivity and upper critical field
measurements~\cite{Kasahara2012} where the mass enhancements are
estimated to be smaller by a factor of $\sim$2 compared to LiFeAs;
this corresponds to a mass enhancement of 1.5--2 over the LDA value. De~Haas-van~Alphen
experiments~\cite{Putzke2012} give values 1.6--3.3 for the mass enhancements.
However, so far no theoretical studies investigating the effects of
correlations on LiFeP have been reported.

Thus, both compounds are considered rather weakly correlated. However,
in the present work 
we argue that the inclusion of correlations has a profound impact on
the Fermi surface topology of both materials which in the case of
LaFePO agrees with ARPES~\cite{Lu2008} observations; for LiFeP, ARPES
measurements are not yet reported and the available dHvA\cite{Putzke2012} data
do not allow to unambiguously decide on the $k_z$ extension of the Fermi surface sheets
as will be discussed below. The features presented here
have not been touched upon in the reported LDA+DMFT calculations on
LaFePO;~\cite{Skornyakov2010, Yin2011} for LiFeP we present, to our knowledge,
 the first LDA+DMFT study in the literature.

We performed full charge self-consistent calculations following the
scheme from Ref.~\onlinecite{Aichhorn2011} using a combination of
electronic structure calculations in the full potential linearized
augmented plane wave (FLAPW) framework as implemented in
WIEN2k~\cite{Blaha2001} with DMFT. For solving the impurity problem we
employed the hybridization expansion continuous-time quantum Monte
Carlo method~\cite{Werner2006} (CT-HYB) as implemented in the ALPS
code;~\cite{Bauer2011, Gull2011} only density-density terms of the
Hund's coupling were considered. We performed calculations on all
available experimental crystal structures as reported in
Refs.~\onlinecite{Kamihara2006, McQueen2008} (LaFePO) and
Refs.~\onlinecite{Deng2009, Putzke2012} (LiFeP) with space group
$P\,4/nmm$; results are shown for the structures from
Ref.~\onlinecite{Kamihara2006} (LaFePO) and Ref.~\onlinecite{Deng2009}
(LiFeP). The energy window for the construction of a localized Wannier
basis was chosen to range from -5.4 eV to 2.7 eV (-6 eV to 3.15 eV)
with respect to the Fermi energy for LaFePO (LiFeP).
The Monte Carlo sampling was done at an inverse temperature $\beta=40$~eV$^{-1}$ with $3 \times 10^6$ sweeps.

For the interaction parameters, we use the definitions of $U=F^0$ and
$J=(F^2+F^4)/14$ in terms of Slater integrals~\cite{Anisimov1997}
$F^k$ with $U=4$~eV, $J=0.8$ eV, and the
around-mean-field~\cite{Czyzyk1994} (AMF) double counting
correction. Some low energy features, in particular the size and shape
of the Fermi hole pockets which are of central interest here, turn out
to be rather sensitive to details of the LDA+DMFT calculation like the
choice of double counting; however, we tested variations in these
technical aspects and found the reported features --while being
affected quantitatively-- qualitatively consistent among calculations
with different double countings, interaction parameters, and reported
crystal structures.

In the following, orbital characters are labeled in a coordinate
system which is 45$^{\circ}$ rotated with respect to the
crystallographic axes, {\it i.e.} $x$ and $y$ point to nearest Fe
neighbors in the Fe-P plane.


\medskip

In Table~\ref{tab:meff} we list the orbital-resolved mass
enhancements for both compounds. The mass enhancements for LaFePO are
in line with the measured values $\sim$1.7\hbox{--}2.2 from the various
experiments~\cite{Lu2008,Qazilbash2009,Coldea2008,Suzuki2009} as well
as previous LDA+DMFT studies where mass enhancements $\sim$1.6\hbox{--}2.2
were calculated.~\cite{Skornyakov2010,Yin2011} Note that for LaFePO
the effective masses are higher for the $e_g$ orbitals whereas in
LiFeP (and most other iron pnictides) the $t_{2g}$ orbitals show
stronger renormalization. This is because of the crystal field
splitting which in LaFePO puts the $t_{2g}$ orbitals below the $e_g$
orbitals thereby promoting a ground state with configuration
$e_g^2t_{2g}^4$ in the atomic limit. This suppresses inter-orbital
fluctuations among the $e_g$ states, rendering these orbitals more
correlated in effect.~\cite{Medici2011,Yin2011} As a consequence, the
$d_{z^2}$ orbital is the most strongly correlated one in LaFePO.

The values for LiFeP range between 1.4 and 1.7 which is roughly a
factor of 2 smaller than in LiFeAs where ARPES~\cite{Borisenko2010}
and dHvA~\cite{Putzke2012} experiments yield mass enhancements of
3--4.  This is in agreement with resistivity and upper critical field
measurements~\cite{Kasahara2012} which also give a factor 2 reduction
with respect to LiFeAs.

\begin{table}[htb]
  \caption{\label{tab:meff} Orbital-resolved mass enhancements $m^\ast/m_{\rm LDA}$.}
\begin{ruledtabular}
\begin{tabular}{llrrrr}
Orbital & & $d_{z^2}$ & $d_{x^2-y^2}$ & $d_{xy}$ & $d_{xz/yz}$ \\
\hline
LaFePO: & $\frac{m^\ast}{m_{\rm LDA}}$ & 1.85 & 1.70 & 1.54 & 1.69\\
LiFeP: & $\frac{m^\ast}{m_{\rm LDA}}$ & 1.52 & 1.39 & 1.71 & 1.62 
\end{tabular}
\end{ruledtabular}
\end{table}

\begin{figure}[htb]
\includegraphics[width=\columnwidth]{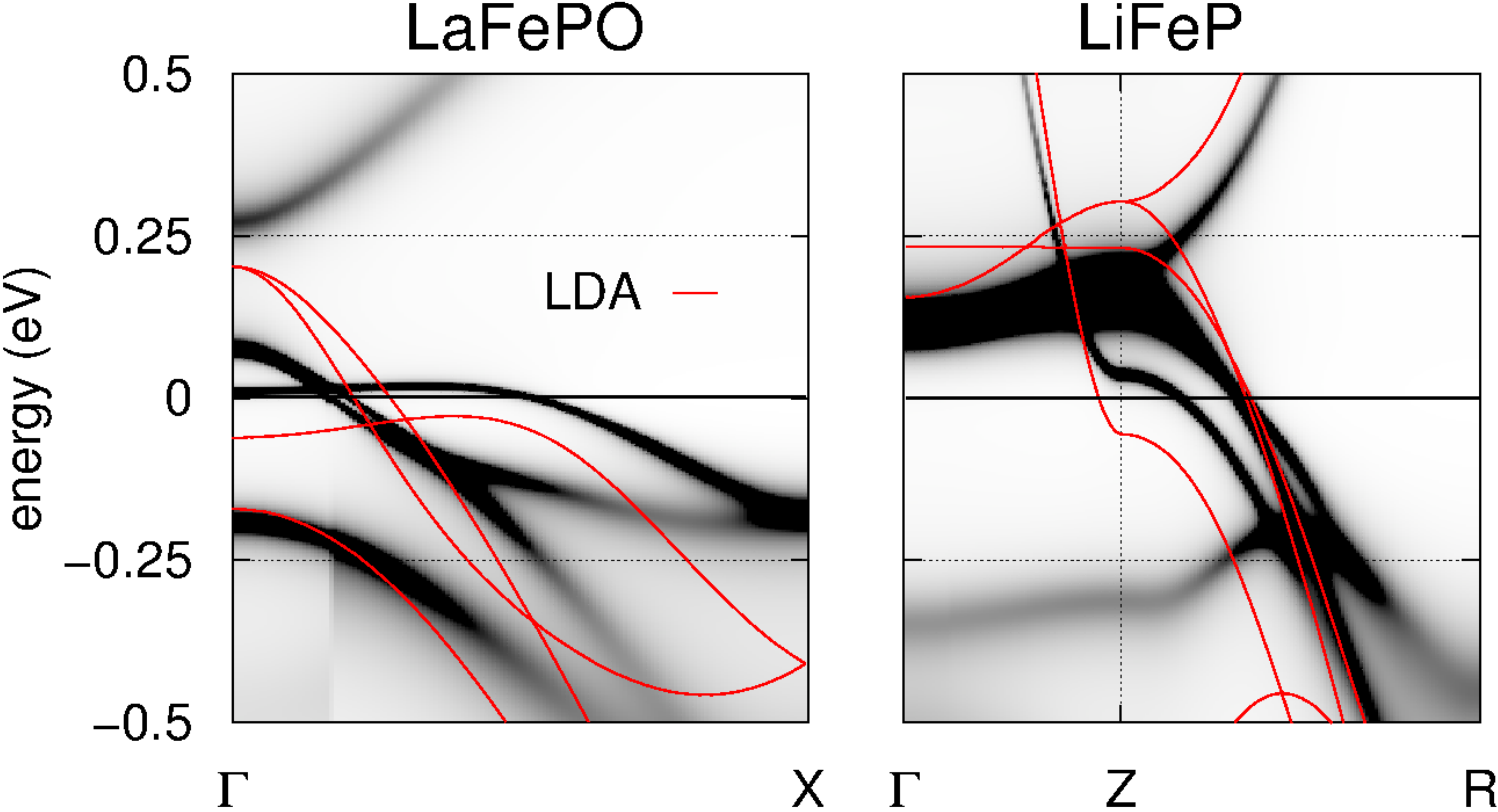}
\caption{\label{fig:Ak}(Color online) Momentum-resolved spectral
  function $A(\bf{k},\omega)$ together with LDA bands in the vicinity of the Fermi surface
topology change.}
\end{figure}

The momentum-resolved spectral function of the two materials is presented in
Fig.~\ref{fig:Ak} in comparison with the LDA band energies. The excitations around the Fermi energy are
well-defined revealing the Fermi liquid nature in accordance with resistivity
measurements.~\cite{Kasahara2012} Most importantly, both compounds feature a band placed just
below the Fermi energy in LDA which gets shifted above $E_F$ upon inclusion of correlations in the
vicinity of the $\Gamma$ (Z) point in LaFePO (LiFeP). In both compounds this band has $d_{z^2}$ orbital
character around $E_F$ for the path shown in Fig.~\ref{fig:Ak} and originates from the
hybridization with phosphorus $p$ states.

\begin{figure*}[htb]
\includegraphics[width=0.9\textwidth]{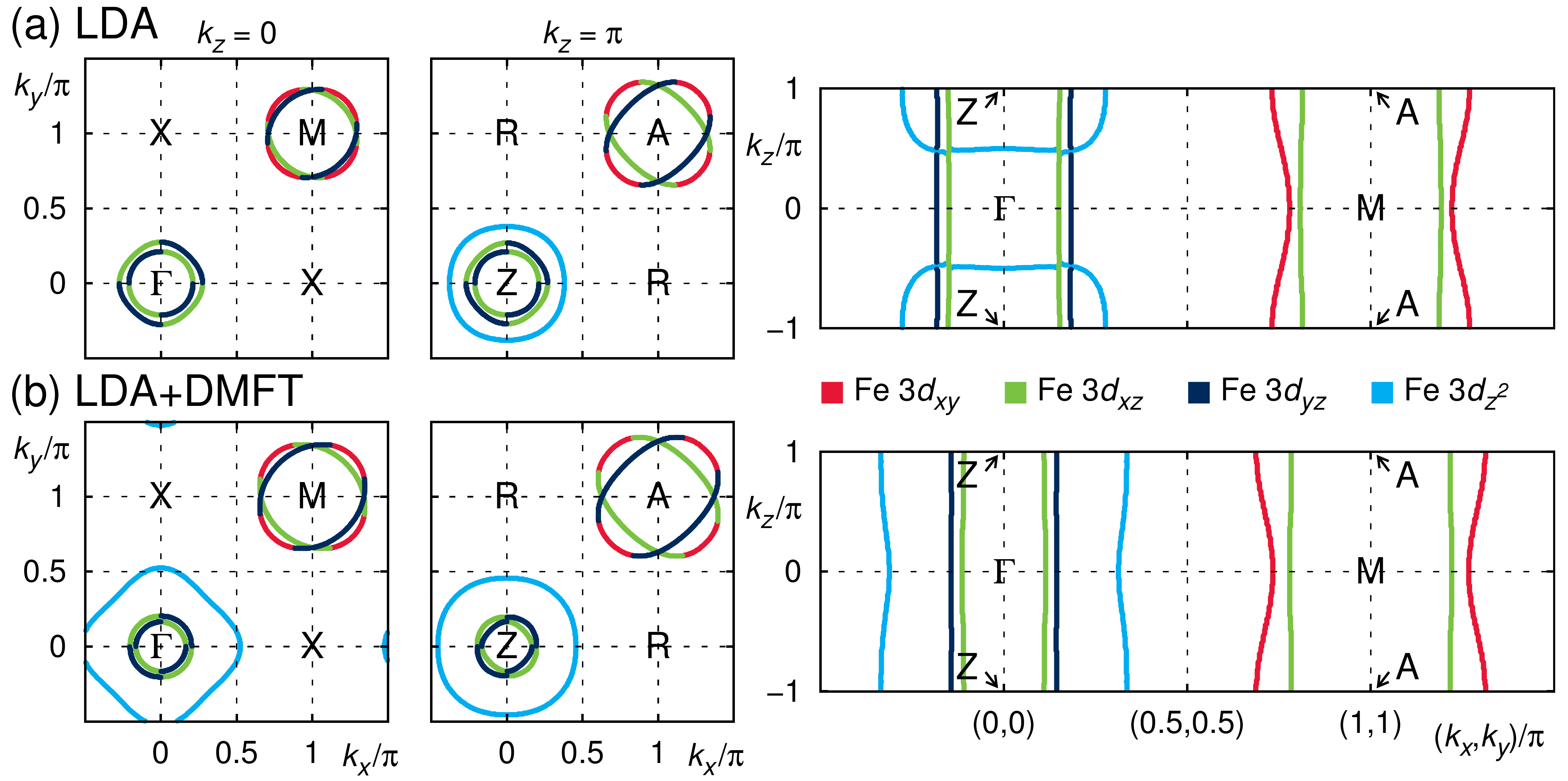}
\caption{\label{fig:lafepo_fermi}(Color online) Fermi surfaces of LaFePO in the $k_z=0$ and
$k_z=\pi$ plane (left panels) and the $k_x=k_y$ plane (right panels) for (a) LDA and (b) LDA+DMFT. The colors give the dominating orbital character.}
\end{figure*}

\begin{figure*}[htb]
\includegraphics[width=0.85\textwidth]{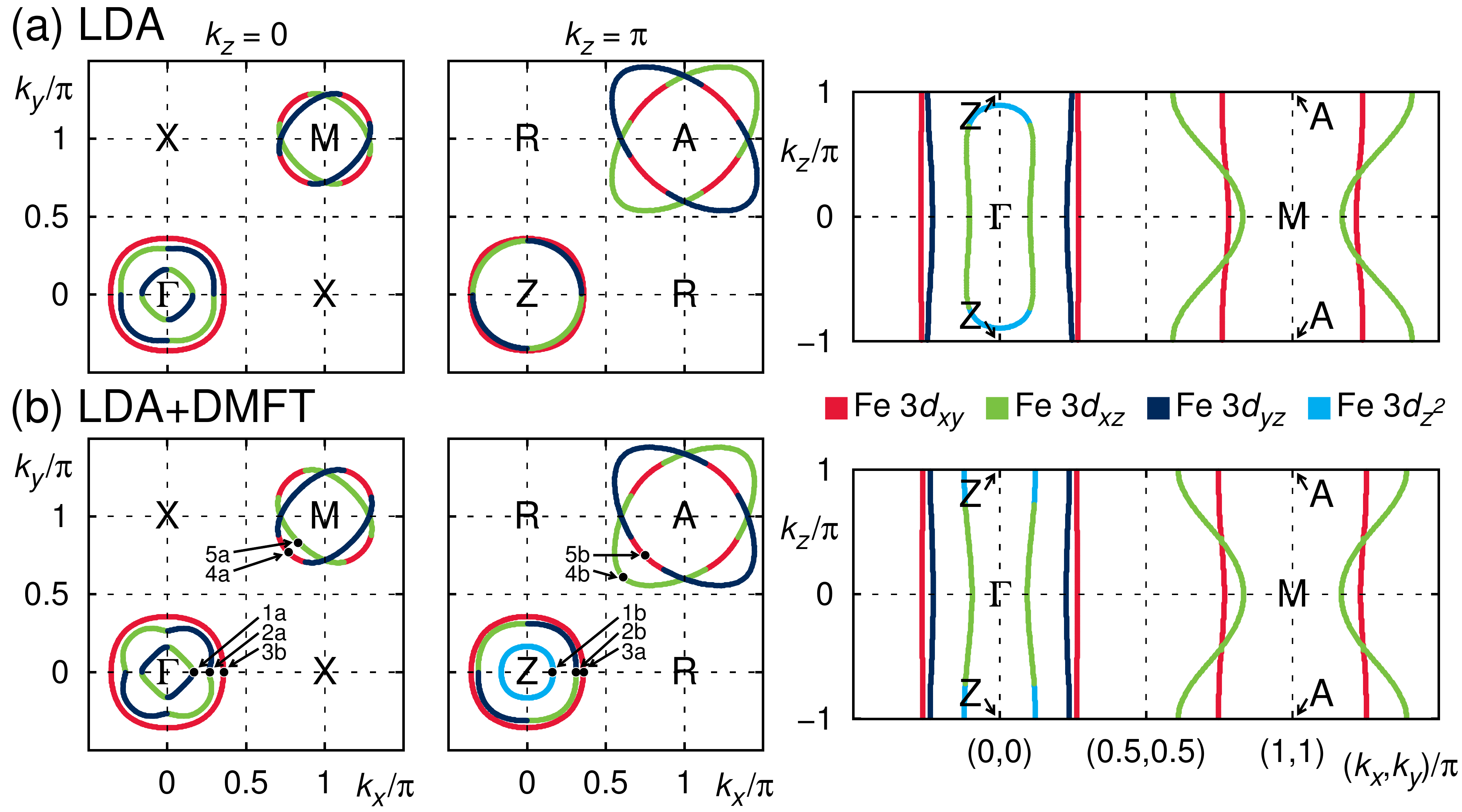}
\caption{\label{fig:lifep_fermi}(Color online) Fermi surfaces of LiFeP in the $k_z=0$ and
$k_z=\pi$ plane
(left panels) and the $k_x=k_y$ plane (rights panels) for (a) LDA and (b) LDA+DMFT. The colors give the dominating orbital character. The arrows indicate the
$k$-points for the calculation of the effective masses in Tab.~\ref{tab:meff_dhva}.}
\end{figure*}

As shown in Figs.~\ref{fig:lafepo_fermi} and \ref{fig:lifep_fermi}, this crossing of the Fermi energy is naturally accompanied by the
appearance at the Fermi surface of an outer hole pocket centered
at $\Gamma$ in LaFePO (see Fig.~\ref{fig:lafepo_fermi}(b))
and an inner hole pocket centered at $Z$ in LiFeP
(see Fig.~\ref{fig:lifep_fermi}(b)). As has been suggested\cite{Kemper2010} for the iron pnictides,
appearance of a pocket with $d_{z^2}$ character may alter the superconducting pairing function to a nodal state and reduce the strength of the pairing as it
is observed in LaFePO and LiFeP in contrast to their arsenic counterparts LaFeAsO and LiFeAs.

\begin{figure}[htb]
\includegraphics[width=\columnwidth]{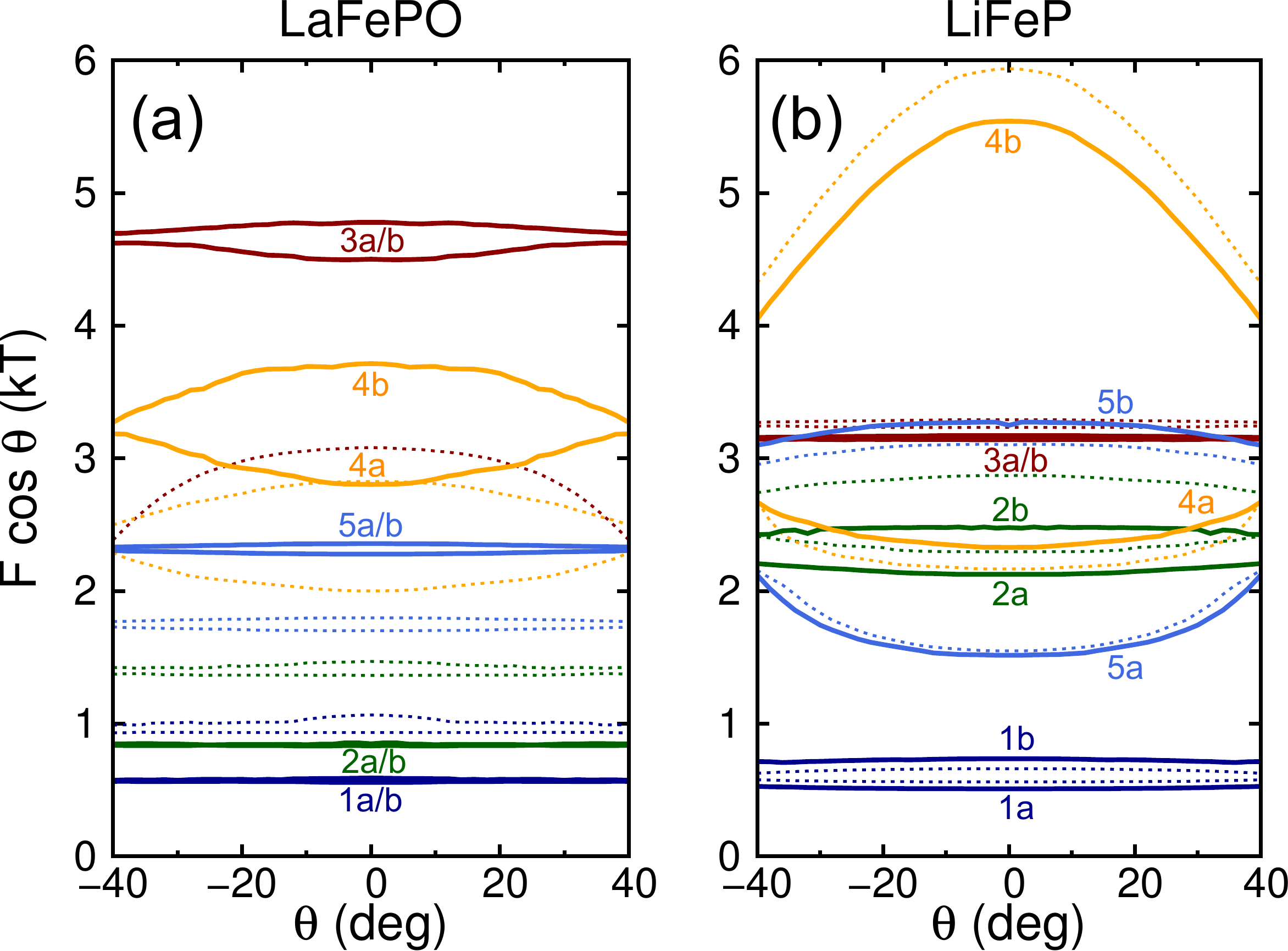}
\caption{\label{fig:dhva}(Color online) dHvA frequencies with respect
  to magnetic field angle obtained within LDA (dashed thin lines) and LDA+DMFT (solid lines). Orbits
1, 2, and 3 refer to the inner, middle, and outer hole pocket, and orbits 4 and 5 to
the outer and inner electron pocket.}
\end{figure}

Due to the hybridization with the phosphorus
states, the position of the crossing band in LDA is sensitive to the
phosphorus $z$ position and doping. Therefore, although the large
outer hole pocket appearing in LaFePO has been observed in
ARPES~\cite{Lu2008}, it has been suspected~\cite{Lu2008,Coldea2008}
that the appearance of this pocket is caused by surface doping
indicated by a too small electron count obtained in ARPES. In
contrast, our calculations yield the change in the Fermi surface
topology as a result of electronic correlations only. Note that the
total charge in the crystal is conserved in our calculations and the
opening of the pocket (i.e. increase of the Fermi surface volume)
merely amounts to a charge transfer from the $d_{z^2}$ orbitals to the
$t_{2g}$ orbitals. Despite the sensitivity of the band position in LDA
(for LaFePO the band energies differ by approx. 12~meV between the two
published structures~\cite{Kamihara2006,McQueen2008}) we found the
opening of the pocket in both structures and with very similar pocket
sizes. As for the electron deficiency in ARPES, the huge size of the measured pocket ($>$12~kT
as compared to $<$5~kT in our calculations, cf. orbit 3a in Fig.~\ref{fig:dhva}(a)) probably still results from a
charge effect.

As a result, the calculated dHvA frequencies for LaFePO experience significant shifts as shown in
Fig.~\ref{fig:dhva}(a). The frequencies correspond to extremal pocket sizes (orbits) that are
observed at a given angle $\Theta$ with respect to the $k_z$ axis. The outer hole pocket
experiences a large increase compared to LDA and the opening at $Z$ adds a new frequency 3a for the
minimal orbit. As a result of charge conservation, the outer electron pocket 4a/b is also blown up.
The enlargement of the electron pocket seen in our calculations is not observed in the dHvA experiment.~\cite{Coldea2008} The
hole pocket itself is not measured in dHvA (7 out of the 10 predicted frequencies are present in the
measurements). In LDA, inclusion of spin-orbit (SO) coupling reduces the size of the inner hole pocket
1a/b but SO coupling is not included in our LDA+DMFT calculations. It is therefore likely that this pocket
shrinks even more than predicted by us, thereby reducing the total Fermi surface volume enclosed by
the hole pockets; this could approximately compensate for the added volume from the opened hole pocket without enlargement of the electron pocket.

For LiFeP, the inclusion of interactions induces only moderate
changes in the sizes of the Fermi surface sheets, see Fig.~\ref{fig:dhva}(b). The frequency shifts with respect to LDA are in
qualitative agreement with the experimental dHvA data from Ref.~\onlinecite{Putzke2012}: the middle
hole pocket 2a/b shifts most and shrinks in size by approx. 0.39~kT (compared to approx. 0.95~kT
in the experiment), the other sheets are less affected (orbit 4b also shrinks substantially but
it it not measured in dHvA). The opening of the inner hole
pocket in LDA+DMFT has almost no effect on the dHvA frequencies: also in LDA, two frequencies for
the inner hole pocket are expected due to a weak peanut-like distortion which gives a minimal
orbit around $\Gamma$ and a maximal orbit at $k_z \approx \nicefrac{\pi}{2}$. In LDA+DMFT, the size
of the pocket at $Z$ (the new maximal orbit) essentially equals the maximal orbit size from LDA,
thereby mimicking the LDA orbits. Since the position of the orbits in the BZ cannot be determined
from dHvA, no clear distinction between the predictions from
LDA and LDA+DMFT can be made from the published data. Measurements up to
$\Theta=90^\circ$ which in principle allow to differentiate between open and closed pockets
($F\cos\theta$ drops to zero for a closed pocket) have been performed and indicate a
tendency of $F\cos\theta$ towards small values
 but are not conclusive because of the very weak signal.~\cite{Note1}

While a good qualitative agreement is reached between our
results and dHvA observations, our calculations do not
 lead to a sufficient shift to attain
complete agreement with the experimental 
frequencies in LiFeP, in particular the calculated reduction
of the middle hole pocket is not pronounced enough. Note that spin-orbit
coupling helps with the size reduction of this pocket, its
effect is comparatively small here, though (about 0.2~kT). Limitations of our approach are also
revealed by a comparison of the effective masses in LiFeP. The effective masses
obtained from the dHvA measurements are rather uniform among all orbits
except for the orbits 2a/b which show only half the mass enhancement of the other orbits. Since
these mass enhancements refer to the Fermi surface orbits rather than the localized orbitals, we
calculated
the LDA+DMFT mass enhancements in the same basis by projecting the self energy in the
localized Wannier basis $\ket{\chi_m(\mathbf{k})}$, $\Sigma_{mm}$, to the basis of Bloch states
$\ket{\psi_\nu(\mathbf{k})}$, $\Sigma_{\nu\nu'}(\mathbf{k})$,
\begin{equation}
 \Sigma_{\nu\nu'}(\mathbf{k}) = \sum_{mm'} P^\ast_{\nu m}(\mathbf k)\;\Sigma_{mm'}\;
P_{m'\nu'}(\mathbf k),
\end{equation}
where $P_{m\nu}(\mathbf k) = \langle \chi_m(\mathbf k) |
\psi_\nu(\mathbf{k}) \rangle$. From the diagonal elements
$\Sigma_{\nu\nu}$ we obtain the mass enhancements of the respective
Fermi surface pocket at the $k$-points indicated in
Fig.~\ref{fig:lifep_fermi}; the values are given in Table~\ref{tab:meff_dhva}.

\begin{table}[htb]
  \caption{
    Mass enhancements of the maximal/minimal Fermi surface orbits in LiFeP. The mass enhancements
are measured at the $k$-points indicated in
Fig.~\ref{fig:lifep_fermi}.}\label{tab:meff_dhva}
\begin{ruledtabular}
\begin{tabular}{lccccc}
orbit & 1a/b & 2a/b & 3a/b & 4a/b & 5a/b \\\hline
pocket & inner & middle & outer& outer & inner\\
&hole& hole&hole& electron&electron\\
\hline
$\frac{m^\ast}{m_{\rm LDA}}$ & 1.47/1.34 & 1.48/1.47 & 1.69/1.70 & 1.49/1.46 & 1.52/1.37
\end{tabular}
\end{ruledtabular}
\end{table}

The significantly smaller mass enhancements of the middle hole pocket
(orbit 2) measured in the quantum oscillation experiments are not seen in
LDA+DMFT. This suggests that this pocket is
differently affected by the coupling to some scattering channel like
spin fluctuations or nonlocal correlations which are not captured by our LDA+DMFT approach.

In summary, we reported LDA+DMFT calculations on LiFeP and LaFePO where we find a change of the
Fermi surface topology upon inclusion of correlations in both compounds, 
namely the opening of an
 outer hole pocket at $\Gamma$ in LaFePO and the opening of an inner hole
pocket at $Z$ in LiFeP, both with Fe~$d_{z^2}$ orbital character.
As discussed by Kemper et al.~\cite{Kemper2010}, this might promote the nodal gap and weak pairing strength, i.e. low $T_c$, in these materials. Whereas this pocket has been observed in
ARPES\cite{Lu2008} for LaFePO, the current experimental situation for LiFeP doesn't allow for a
conclusive testing of our predictions and further experimental work is desired. Also we find
that the pecularities of the middle hole pocket in LiFeP observed in dHvA
experiments and not reproduced  in our LDA+DMFT approach
 reveal the importance
of scattering channels beyond local correlations.  
 
We would like to thank A. Coldea, C. Putzke, A. Carrington, and P. J. Hirschfeld for
useful discussions and
the Deutsche Forschungsgemeinschaft for financial support through
grant SPP 1458, the Helmholtz Association for support through
HA216/EMMI and the centre for scientific computing (CSC,LOEWE-CSC) in Frankfurt
for computing facilities.


\begin{thebibliography}{30}%
\makeatletter
\providecommand \@ifxundefined [1]{%
 \@ifx{#1\undefined}
}%
\providecommand \@ifnum [1]{%
 \ifnum #1\expandafter \@firstoftwo
 \else \expandafter \@secondoftwo
 \fi
}%
\providecommand \@ifx [1]{%
 \ifx #1\expandafter \@firstoftwo
 \else \expandafter \@secondoftwo
 \fi
}%
\providecommand \natexlab [1]{#1}%
\providecommand \enquote  [1]{``#1''}%
\providecommand \bibnamefont  [1]{#1}%
\providecommand \bibfnamefont [1]{#1}%
\providecommand \citenamefont [1]{#1}%
\providecommand \href@noop [0]{\@secondoftwo}%
\providecommand \href [0]{\begingroup \@sanitize@url \@href}%
\providecommand \@href[1]{\@@startlink{#1}\@@href}%
\providecommand \@@href[1]{\endgroup#1\@@endlink}%
\providecommand \@sanitize@url [0]{\catcode `\\12\catcode `\$12\catcode
  `\&12\catcode `\#12\catcode `\^12\catcode `\_12\catcode `\%12\relax}%
\providecommand \@@startlink[1]{}%
\providecommand \@@endlink[0]{}%
\providecommand \url  [0]{\begingroup\@sanitize@url \@url }%
\providecommand \@url [1]{\endgroup\@href {#1}{\urlprefix }}%
\providecommand \urlprefix  [0]{URL }%
\providecommand \Eprint [0]{\href }%
\providecommand \doibase [0]{http://dx.doi.org/}%
\providecommand \selectlanguage [0]{\@gobble}%
\providecommand \bibinfo  [0]{\@secondoftwo}%
\providecommand \bibfield  [0]{\@secondoftwo}%
\providecommand \translation [1]{[#1]}%
\providecommand \BibitemOpen [0]{}%
\providecommand \bibitemStop [0]{}%
\providecommand \bibitemNoStop [0]{.\EOS\space}%
\providecommand \EOS [0]{\spacefactor3000\relax}%
\providecommand \BibitemShut  [1]{\csname bibitem#1\endcsname}%
\let\auto@bib@innerbib\@empty
\bibitem [{\citenamefont {Kemper}\ \emph {et~al.}(2010)\citenamefont {Kemper},
  \citenamefont {Maier}, \citenamefont {Graser}, \citenamefont {Cheng},
  \citenamefont {Hirschfeld},\ and\ \citenamefont {Scalapino}}]{Kemper2010}%
  \BibitemOpen
  \bibfield  {author} {\bibinfo {author} {\bibfnamefont {A.~F.}\ \bibnamefont
  {Kemper}}, \bibinfo {author} {\bibfnamefont {T.~A.}\ \bibnamefont {Maier}},
  \bibinfo {author} {\bibfnamefont {S.}~\bibnamefont {Graser}}, \bibinfo
  {author} {\bibfnamefont {H.-P.}\ \bibnamefont {Cheng}}, \bibinfo {author}
  {\bibfnamefont {P.~J.}\ \bibnamefont {Hirschfeld}}, \ and\ \bibinfo {author}
  {\bibfnamefont {D.~J.}\ \bibnamefont {Scalapino}},\ }\href@noop {} {\bibfield
   {journal} {\bibinfo  {journal} {New J. Phys.}\ }\textbf {\bibinfo
  {volume} {12}},\ \bibinfo {pages} {073030} (\bibinfo {year}
  {2010})}\BibitemShut {NoStop}%
\bibitem [{\citenamefont {Wu}\ \emph {et~al.}(2009)\citenamefont {Wu},
  \citenamefont {Xie}, \citenamefont {Chen}, \citenamefont {Zhong},
  \citenamefont {Liu}, \citenamefont {Shi}, \citenamefont {Li}, \citenamefont
  {Wang}, \citenamefont {Wu}, \citenamefont {Yan}, \citenamefont {Ying},\ and\
  \citenamefont {Chen}}]{Wu2009}%
  \BibitemOpen
  \bibfield  {author} {\bibinfo {author} {\bibfnamefont {G.}~\bibnamefont
  {Wu}}, \bibinfo {author} {\bibfnamefont {Y.~L.}\ \bibnamefont {Xie}},
  \bibinfo {author} {\bibfnamefont {H.}~\bibnamefont {Chen}}, \bibinfo {author}
  {\bibfnamefont {M.}~\bibnamefont {Zhong}}, \bibinfo {author} {\bibfnamefont
  {R.~H.}\ \bibnamefont {Liu}}, \bibinfo {author} {\bibfnamefont {B.~C.}\
  \bibnamefont {Shi}}, \bibinfo {author} {\bibfnamefont {Q.~J.}\ \bibnamefont
  {Li}}, \bibinfo {author} {\bibfnamefont {X.~F.}\ \bibnamefont {Wang}},
  \bibinfo {author} {\bibfnamefont {T.}~\bibnamefont {Wu}}, \bibinfo {author}
  {\bibfnamefont {Y.~J.}\ \bibnamefont {Yan}}, \bibinfo {author} {\bibfnamefont
  {J.~J.}\ \bibnamefont {Ying}}, \ and\ \bibinfo {author} {\bibfnamefont
  {X.~H.}\ \bibnamefont {Chen}},\ }\href@noop {} {\bibfield  {journal}
  {\bibinfo  {journal} {Journal of Physics: Condensed Matter}\ }\textbf
  {\bibinfo {volume} {21}},\ \bibinfo {pages} {142203} (\bibinfo {year}
  {2009})}\BibitemShut {NoStop}%
\bibitem [{\citenamefont {Kamihara}\ \emph {et~al.}(2006)\citenamefont
  {Kamihara}, \citenamefont {Hiramatsu}, \citenamefont {Hirano}, \citenamefont
  {Kawamura}, \citenamefont {Yanagi}, \citenamefont {Kamiya},\ and\
  \citenamefont {Hosono}}]{Kamihara2006}%
  \BibitemOpen
  \bibfield  {author} {\bibinfo {author} {\bibfnamefont {Y.}~\bibnamefont
  {Kamihara}}, \bibinfo {author} {\bibfnamefont {H.}~\bibnamefont {Hiramatsu}},
  \bibinfo {author} {\bibfnamefont {M.}~\bibnamefont {Hirano}}, \bibinfo
  {author} {\bibfnamefont {R.}~\bibnamefont {Kawamura}}, \bibinfo {author}
  {\bibfnamefont {H.}~\bibnamefont {Yanagi}}, \bibinfo {author} {\bibfnamefont
  {T.}~\bibnamefont {Kamiya}}, \ and\ \bibinfo {author} {\bibfnamefont
  {H.}~\bibnamefont {Hosono}},\ }\href@noop {} {\bibfield  {journal} {\bibinfo
  {journal} {J. Am. Chem. Soc.}\ }\textbf {\bibinfo {volume} {128}},\ \bibinfo
  {pages} {10012} (\bibinfo {year} {2006})}\BibitemShut {NoStop}%
\bibitem [{\citenamefont {McQueen}\ \emph {et~al.}(2008)\citenamefont
  {McQueen}, \citenamefont {Regulacio}, \citenamefont {Williams}, \citenamefont
  {Huang}, \citenamefont {Lynn}, \citenamefont {Hor}, \citenamefont {West},
  \citenamefont {Green},\ and\ \citenamefont {Cava}}]{McQueen2008}%
  \BibitemOpen
  \bibfield  {author} {\bibinfo {author} {\bibfnamefont {T.~M.}\ \bibnamefont
  {McQueen}}, \bibinfo {author} {\bibfnamefont {M.}~\bibnamefont {Regulacio}},
  \bibinfo {author} {\bibfnamefont {A.~J.}\ \bibnamefont {Williams}}, \bibinfo
  {author} {\bibfnamefont {Q.}~\bibnamefont {Huang}}, \bibinfo {author}
  {\bibfnamefont {J.~W.}\ \bibnamefont {Lynn}}, \bibinfo {author}
  {\bibfnamefont {Y.~S.}\ \bibnamefont {Hor}}, \bibinfo {author} {\bibfnamefont
  {D.~V.}\ \bibnamefont {West}}, \bibinfo {author} {\bibfnamefont {M.~A.}\
  \bibnamefont {Green}}, \ and\ \bibinfo {author} {\bibfnamefont {R.~J.}\
  \bibnamefont {Cava}},\ }\href@noop {} {\bibfield  {journal} {\bibinfo
  {journal} {Phys. Rev. B}\ }\textbf {\bibinfo {volume} {78}},\ \bibinfo
  {pages} {024521} (\bibinfo {year} {2008})}\BibitemShut {NoStop}%
\bibitem [{\citenamefont {Yamashita}\ \emph {et~al.}(2009)\citenamefont
  {Yamashita}, \citenamefont {Nakata}, \citenamefont {Senshu}, \citenamefont
  {Tonegawa}, \citenamefont {Ikada}, \citenamefont {Hashimoto}, \citenamefont
  {Sugawara}, \citenamefont {Shibauchi},\ and\ \citenamefont
  {Matsuda}}]{Yamashita2009}%
  \BibitemOpen
  \bibfield  {author} {\bibinfo {author} {\bibfnamefont {M.}~\bibnamefont
  {Yamashita}}, \bibinfo {author} {\bibfnamefont {N.}~\bibnamefont {Nakata}},
  \bibinfo {author} {\bibfnamefont {Y.}~\bibnamefont {Senshu}}, \bibinfo
  {author} {\bibfnamefont {S.}~\bibnamefont {Tonegawa}}, \bibinfo {author}
  {\bibfnamefont {K.}~\bibnamefont {Ikada}}, \bibinfo {author} {\bibfnamefont
  {K.}~\bibnamefont {Hashimoto}}, \bibinfo {author} {\bibfnamefont
  {H.}~\bibnamefont {Sugawara}}, \bibinfo {author} {\bibfnamefont
  {T.}~\bibnamefont {Shibauchi}}, \ and\ \bibinfo {author} {\bibfnamefont
  {Y.}~\bibnamefont {Matsuda}},\ }\href@noop {} {\bibfield  {journal} {\bibinfo
   {journal} {Phys. Rev. B}\ }\textbf {\bibinfo {volume} {80}},\ \bibinfo
  {pages} {220509} (\bibinfo {year} {2009})}\BibitemShut {NoStop}%
\bibitem [{\citenamefont {Fletcher}\ \emph {et~al.}(2009)\citenamefont
  {Fletcher}, \citenamefont {Serafin}, \citenamefont {Malone}, \citenamefont
  {Analytis}, \citenamefont {Chu}, \citenamefont {Erickson}, \citenamefont
  {Fisher},\ and\ \citenamefont {Carrington}}]{Fletcher2009}%
  \BibitemOpen
  \bibfield  {author} {\bibinfo {author} {\bibfnamefont {J.~D.}\ \bibnamefont
  {Fletcher}}, \bibinfo {author} {\bibfnamefont {A.}~\bibnamefont {Serafin}},
  \bibinfo {author} {\bibfnamefont {L.}~\bibnamefont {Malone}}, \bibinfo
  {author} {\bibfnamefont {J.~G.}\ \bibnamefont {Analytis}}, \bibinfo {author}
  {\bibfnamefont {J.-H.}\ \bibnamefont {Chu}}, \bibinfo {author} {\bibfnamefont
  {A.~S.}\ \bibnamefont {Erickson}}, \bibinfo {author} {\bibfnamefont {I.~R.}\
  \bibnamefont {Fisher}}, \ and\ \bibinfo {author} {\bibfnamefont
  {A.}~\bibnamefont {Carrington}},\ }\href@noop {} {\bibfield  {journal}
  {\bibinfo  {journal} {Phys. Rev. Lett.}\ }\textbf {\bibinfo {volume} {102}},\
  \bibinfo {pages} {147001} (\bibinfo {year} {2009})}\BibitemShut {NoStop}%
\bibitem [{\citenamefont {Hicks}\ \emph {et~al.}(2009)\citenamefont {Hicks},
  \citenamefont {Lippman}, \citenamefont {Huber}, \citenamefont {Analytis},
  \citenamefont {Chu}, \citenamefont {Erickson}, \citenamefont {Fisher},\ and\
  \citenamefont {Moler}}]{Hicks2009}%
  \BibitemOpen
  \bibfield  {author} {\bibinfo {author} {\bibfnamefont {C.~W.}\ \bibnamefont
  {Hicks}}, \bibinfo {author} {\bibfnamefont {T.~M.}\ \bibnamefont {Lippman}},
  \bibinfo {author} {\bibfnamefont {M.~E.}\ \bibnamefont {Huber}}, \bibinfo
  {author} {\bibfnamefont {J.~G.}\ \bibnamefont {Analytis}}, \bibinfo {author}
  {\bibfnamefont {J.-H.}\ \bibnamefont {Chu}}, \bibinfo {author} {\bibfnamefont
  {A.~S.}\ \bibnamefont {Erickson}}, \bibinfo {author} {\bibfnamefont {I.~R.}\
  \bibnamefont {Fisher}}, \ and\ \bibinfo {author} {\bibfnamefont {K.~A.}\
  \bibnamefont {Moler}},\ }\href@noop {} {\bibfield  {journal} {\bibinfo
  {journal} {Phys. Rev. Lett.}\ }\textbf {\bibinfo {volume} {103}},\ \bibinfo
  {pages} {127003} (\bibinfo {year} {2009})}\BibitemShut {NoStop}%
\bibitem [{\citenamefont {Lu}\ \emph {et~al.}(2008)\citenamefont {Lu},
  \citenamefont {Yi}, \citenamefont {Mo}, \citenamefont {Erickson},
  \citenamefont {Analytis}, \citenamefont {Chu}, \citenamefont {Singh},
  \citenamefont {Hussain}, \citenamefont {Geballe}, \citenamefont {Fisher},\
  and\ \citenamefont {Shen}}]{Lu2008}%
  \BibitemOpen
  \bibfield  {author} {\bibinfo {author} {\bibfnamefont {D.~H.}\ \bibnamefont
  {Lu}}, \bibinfo {author} {\bibfnamefont {M.}~\bibnamefont {Yi}}, \bibinfo
  {author} {\bibfnamefont {S.-K.}\ \bibnamefont {Mo}}, \bibinfo {author}
  {\bibfnamefont {A.~S.}\ \bibnamefont {Erickson}}, \bibinfo {author}
  {\bibfnamefont {J.}~\bibnamefont {Analytis}}, \bibinfo {author}
  {\bibfnamefont {J.-H.}\ \bibnamefont {Chu}}, \bibinfo {author} {\bibfnamefont
  {D.~J.}\ \bibnamefont {Singh}}, \bibinfo {author} {\bibfnamefont
  {Z.}~\bibnamefont {Hussain}}, \bibinfo {author} {\bibfnamefont {T.~H.}\
  \bibnamefont {Geballe}}, \bibinfo {author} {\bibfnamefont {I.~R.}\
  \bibnamefont {Fisher}}, \ and\ \bibinfo {author} {\bibfnamefont {Z.-X.}\
  \bibnamefont {Shen}},\ }\href@noop {} {\bibfield  {journal} {\bibinfo
  {journal} {Nature}\ }\textbf {\bibinfo {volume} {455}},\ \bibinfo {pages}
  {81} (\bibinfo {year} {2008})}\BibitemShut {NoStop}%
\bibitem [{\citenamefont {Qazilbash}\ \emph {et~al.}(2009)\citenamefont
  {Qazilbash}, \citenamefont {Hamlin}, \citenamefont {Baumbach}, \citenamefont
  {Zhang}, \citenamefont {Singh}, \citenamefont {Maple},\ and\ \citenamefont
  {Basov}}]{Qazilbash2009}%
  \BibitemOpen
  \bibfield  {author} {\bibinfo {author} {\bibfnamefont {M.~M.}\ \bibnamefont
  {Qazilbash}}, \bibinfo {author} {\bibfnamefont {J.~J.}\ \bibnamefont
  {Hamlin}}, \bibinfo {author} {\bibfnamefont {R.~E.}\ \bibnamefont
  {Baumbach}}, \bibinfo {author} {\bibfnamefont {L.}~\bibnamefont {Zhang}},
  \bibinfo {author} {\bibfnamefont {D.~J.}\ \bibnamefont {Singh}}, \bibinfo
  {author} {\bibfnamefont {M.~B.}\ \bibnamefont {Maple}}, \ and\ \bibinfo
  {author} {\bibfnamefont {D.~N.}\ \bibnamefont {Basov}},\ }\href@noop {}
  {\bibfield  {journal} {\bibinfo  {journal} {Nat. Phys.}\ }\textbf {\bibinfo
  {volume} {5}},\ \bibinfo {pages} {647} (\bibinfo {year} {2009})}\BibitemShut
  {NoStop}%
\bibitem [{\citenamefont {Coldea}\ \emph {et~al.}(2008)\citenamefont {Coldea},
  \citenamefont {Fletcher}, \citenamefont {Carrington}, \citenamefont
  {Analytis}, \citenamefont {Bangura}, \citenamefont {Chu}, \citenamefont
  {Erickson}, \citenamefont {Fisher}, \citenamefont {Hussey},\ and\
  \citenamefont {McDonald}}]{Coldea2008}%
  \BibitemOpen
  \bibfield  {author} {\bibinfo {author} {\bibfnamefont {A.~I.}\ \bibnamefont
  {Coldea}}, \bibinfo {author} {\bibfnamefont {J.~D.}\ \bibnamefont
  {Fletcher}}, \bibinfo {author} {\bibfnamefont {A.}~\bibnamefont
  {Carrington}}, \bibinfo {author} {\bibfnamefont {J.~G.}\ \bibnamefont
  {Analytis}}, \bibinfo {author} {\bibfnamefont {A.~F.}\ \bibnamefont
  {Bangura}}, \bibinfo {author} {\bibfnamefont {J.-H.}\ \bibnamefont {Chu}},
  \bibinfo {author} {\bibfnamefont {A.~S.}\ \bibnamefont {Erickson}}, \bibinfo
  {author} {\bibfnamefont {I.~R.}\ \bibnamefont {Fisher}}, \bibinfo {author}
  {\bibfnamefont {N.~E.}\ \bibnamefont {Hussey}}, \ and\ \bibinfo {author}
  {\bibfnamefont {R.~D.}\ \bibnamefont {McDonald}},\ }\href@noop {} {\bibfield
  {journal} {\bibinfo  {journal} {Phys. Rev. Lett.}\ }\textbf {\bibinfo
  {volume} {101}},\ \bibinfo {pages} {216402} (\bibinfo {year}
  {2008})}\BibitemShut {NoStop}%
\bibitem [{\citenamefont {Suzuki}\ \emph {et~al.}(2009)\citenamefont {Suzuki},
  \citenamefont {Miyasaka}, \citenamefont {Tajima}, \citenamefont {Kida},\ and\
  \citenamefont {Hagiwara}}]{Suzuki2009}%
  \BibitemOpen
  \bibfield  {author} {\bibinfo {author} {\bibfnamefont {S.}~\bibnamefont
  {Suzuki}}, \bibinfo {author} {\bibfnamefont {S.}~\bibnamefont {Miyasaka}},
  \bibinfo {author} {\bibfnamefont {S.}~\bibnamefont {Tajima}}, \bibinfo
  {author} {\bibfnamefont {T.}~\bibnamefont {Kida}}, \ and\ \bibinfo {author}
  {\bibfnamefont {M.}~\bibnamefont {Hagiwara}},\ }\href@noop {} {\bibfield
  {journal} {\bibinfo  {journal} {J. Phys. Soc. Jpn.}\ }\textbf {\bibinfo
  {volume} {78}},\ \bibinfo {pages} {114712} (\bibinfo {year}
  {2009})}\BibitemShut {NoStop}%
\bibitem [{\citenamefont {Skornyakov}\ \emph {et~al.}(2010)\citenamefont
  {Skornyakov}, \citenamefont {Skorikov}, \citenamefont {Lukoyanov},
  \citenamefont {Shorikov},\ and\ \citenamefont {Anisimov}}]{Skornyakov2010}%
  \BibitemOpen
  \bibfield  {author} {\bibinfo {author} {\bibfnamefont {S.~L.}\ \bibnamefont
  {Skornyakov}}, \bibinfo {author} {\bibfnamefont {N.~A.}\ \bibnamefont
  {Skorikov}}, \bibinfo {author} {\bibfnamefont {A.~V.}\ \bibnamefont
  {Lukoyanov}}, \bibinfo {author} {\bibfnamefont {A.~O.}\ \bibnamefont
  {Shorikov}}, \ and\ \bibinfo {author} {\bibfnamefont {V.~I.}\ \bibnamefont
  {Anisimov}},\ }\href@noop {} {\bibfield  {journal} {\bibinfo  {journal}
  {Phys. Rev. B}\ }\textbf {\bibinfo {volume} {81}},\ \bibinfo {pages} {174522}
  (\bibinfo {year} {2010})}\BibitemShut {NoStop}%
\bibitem [{\citenamefont {Yin}\ \emph {et~al.}(2011)\citenamefont {Yin},
  \citenamefont {Haule},\ and\ \citenamefont {Kotliar}}]{Yin2011}%
  \BibitemOpen
  \bibfield  {author} {\bibinfo {author} {\bibfnamefont {Z.~P.}\ \bibnamefont
  {Yin}}, \bibinfo {author} {\bibfnamefont {K.}~\bibnamefont {Haule}}, \ and\
  \bibinfo {author} {\bibfnamefont {G.}~\bibnamefont {Kotliar}},\ }\href@noop
  {} {\bibfield  {journal} {\bibinfo  {journal} {Nature Mater.}\ }\textbf
  {\bibinfo {volume} {10}},\ \bibinfo {pages} {932} (\bibinfo {year}
  {2011})}\BibitemShut {NoStop}%
\bibitem [{\citenamefont {Tapp}\ \emph {et~al.}(2008)\citenamefont {Tapp},
  \citenamefont {Tang}, \citenamefont {Lv}, \citenamefont {Sasmal},
  \citenamefont {Lorenz}, \citenamefont {Chu},\ and\ \citenamefont
  {Guloy}}]{Tapp2008}%
  \BibitemOpen
  \bibfield  {author} {\bibinfo {author} {\bibfnamefont {J.~H.}\ \bibnamefont
  {Tapp}}, \bibinfo {author} {\bibfnamefont {Z.}~\bibnamefont {Tang}}, \bibinfo
  {author} {\bibfnamefont {B.}~\bibnamefont {Lv}}, \bibinfo {author}
  {\bibfnamefont {K.}~\bibnamefont {Sasmal}}, \bibinfo {author} {\bibfnamefont
  {B.}~\bibnamefont {Lorenz}}, \bibinfo {author} {\bibfnamefont {P.~C.~W.}\
  \bibnamefont {Chu}}, \ and\ \bibinfo {author} {\bibfnamefont {A.~M.}\
  \bibnamefont {Guloy}},\ }\href@noop {} {\bibfield  {journal} {\bibinfo
  {journal} {Phys. Rev. B}\ }\textbf {\bibinfo {volume} {78}},\ \bibinfo
  {pages} {060505} (\bibinfo {year} {2008})}\BibitemShut {NoStop}%
\bibitem [{\citenamefont {Deng}\ \emph {et~al.}(2009)\citenamefont {Deng},
  \citenamefont {Wang}, \citenamefont {Liu}, \citenamefont {Zhang},
  \citenamefont {Lv}, \citenamefont {Zhu}, \citenamefont {Yu},\ and\
  \citenamefont {Jin}}]{Deng2009}%
  \BibitemOpen
  \bibfield  {author} {\bibinfo {author} {\bibfnamefont {Z.}~\bibnamefont
  {Deng}}, \bibinfo {author} {\bibfnamefont {X.~C.}\ \bibnamefont {Wang}},
  \bibinfo {author} {\bibfnamefont {Q.~Q.}\ \bibnamefont {Liu}}, \bibinfo
  {author} {\bibfnamefont {S.~J.}\ \bibnamefont {Zhang}}, \bibinfo {author}
  {\bibfnamefont {Y.~X.}\ \bibnamefont {Lv}}, \bibinfo {author} {\bibfnamefont
  {J.~L.}\ \bibnamefont {Zhu}}, \bibinfo {author} {\bibfnamefont {R.~C.}\
  \bibnamefont {Yu}}, \ and\ \bibinfo {author} {\bibfnamefont {C.~Q.}\
  \bibnamefont {Jin}},\ }\href@noop {} {\bibfield  {journal} {\bibinfo
  {journal} {Europhys. Lett.}\ }\textbf {\bibinfo {volume} {87}},\ \bibinfo
  {pages} {37004} (\bibinfo {year} {2009})}\BibitemShut {NoStop}%
\bibitem [{\citenamefont {Hashimoto}\ \emph {et~al.}(2012)\citenamefont
  {Hashimoto}, \citenamefont {Kasahara}, \citenamefont {Katsumata},
  \citenamefont {Mizukami}, \citenamefont {Yamashita}, \citenamefont {Ikeda},
  \citenamefont {Terashima}, \citenamefont {Carrington}, \citenamefont
  {Matsuda},\ and\ \citenamefont {Shibauchi}}]{Hashimoto2012}%
  \BibitemOpen
  \bibfield  {author} {\bibinfo {author} {\bibfnamefont {K.}~\bibnamefont
  {Hashimoto}}, \bibinfo {author} {\bibfnamefont {S.}~\bibnamefont {Kasahara}},
  \bibinfo {author} {\bibfnamefont {R.}~\bibnamefont {Katsumata}}, \bibinfo
  {author} {\bibfnamefont {Y.}~\bibnamefont {Mizukami}}, \bibinfo {author}
  {\bibfnamefont {M.}~\bibnamefont {Yamashita}}, \bibinfo {author}
  {\bibfnamefont {H.}~\bibnamefont {Ikeda}}, \bibinfo {author} {\bibfnamefont
  {T.}~\bibnamefont {Terashima}}, \bibinfo {author} {\bibfnamefont
  {A.}~\bibnamefont {Carrington}}, \bibinfo {author} {\bibfnamefont
  {Y.}~\bibnamefont {Matsuda}}, \ and\ \bibinfo {author} {\bibfnamefont
  {T.}~\bibnamefont {Shibauchi}},\ }\href@noop {} {\bibfield  {journal}
  {\bibinfo  {journal} {Phys. Rev. Lett.}\ }\textbf {\bibinfo {volume} {108}},\
  \bibinfo {pages} {047003} (\bibinfo {year} {2012})}\BibitemShut {NoStop}%
\bibitem [{\citenamefont {Kuroki}\ \emph {et~al.}(2009)\citenamefont {Kuroki},
  \citenamefont {Usui}, \citenamefont {Onari}, \citenamefont {Arita},\ and\
  \citenamefont {Aoki}}]{Kuroki2009}%
  \BibitemOpen
  \bibfield  {author} {\bibinfo {author} {\bibfnamefont {K.}~\bibnamefont
  {Kuroki}}, \bibinfo {author} {\bibfnamefont {H.}~\bibnamefont {Usui}},
  \bibinfo {author} {\bibfnamefont {S.}~\bibnamefont {Onari}}, \bibinfo
  {author} {\bibfnamefont {R.}~\bibnamefont {Arita}}, \ and\ \bibinfo {author}
  {\bibfnamefont {H.}~\bibnamefont {Aoki}},\ }\href@noop {} {\bibfield
  {journal} {\bibinfo  {journal} {Phys. Rev. B}\ }\textbf {\bibinfo {volume}
  {79}},\ \bibinfo {pages} {224511} (\bibinfo {year} {2009})}\BibitemShut
  {NoStop}%
\bibitem [{\citenamefont {Ikeda}\ \emph {et~al.}(2010)\citenamefont {Ikeda},
  \citenamefont {Arita},\ and\ \citenamefont {Kune\ifmmode~\check{s}\else
  \v{s}\fi{}}}]{Ikeda2010}%
  \BibitemOpen
  \bibfield  {author} {\bibinfo {author} {\bibfnamefont {H.}~\bibnamefont
  {Ikeda}}, \bibinfo {author} {\bibfnamefont {R.}~\bibnamefont {Arita}}, \ and\
  \bibinfo {author} {\bibfnamefont {J.}~\bibnamefont
  {Kune\ifmmode~\check{s}\else \v{s}\fi{}}},\ }\href@noop {} {\bibfield
  {journal} {\bibinfo  {journal} {Phys. Rev. B}\ }\textbf {\bibinfo {volume}
  {81}},\ \bibinfo {pages} {054502} (\bibinfo {year} {2010})}\BibitemShut
  {NoStop}%
\bibitem [{\citenamefont {Thomale}\ \emph {et~al.}(2011)\citenamefont
  {Thomale}, \citenamefont {Platt}, \citenamefont {Hanke},\ and\ \citenamefont
  {Bernevig}}]{Thomale2011}%
  \BibitemOpen
  \bibfield  {author} {\bibinfo {author} {\bibfnamefont {R.}~\bibnamefont
  {Thomale}}, \bibinfo {author} {\bibfnamefont {C.}~\bibnamefont {Platt}},
  \bibinfo {author} {\bibfnamefont {W.}~\bibnamefont {Hanke}}, \ and\ \bibinfo
  {author} {\bibfnamefont {B.~A.}\ \bibnamefont {Bernevig}},\ }\href@noop {}
  {\bibfield  {journal} {\bibinfo  {journal} {Phys. Rev. Lett.}\ }\textbf
  {\bibinfo {volume} {106}},\ \bibinfo {pages} {187003} (\bibinfo {year}
  {2011})}\BibitemShut {NoStop}%
\bibitem [{\citenamefont {Hirschfeld}\ \emph {et~al.}(2011)\citenamefont
  {Hirschfeld}, \citenamefont {Korshunov},\ and\ \citenamefont
  {Mazin}}]{Hirschfeld2011}%
  \BibitemOpen
  For a review see \bibfield  {author} {\bibinfo {author} {\bibfnamefont {P.~J.}\ \bibnamefont
  {Hirschfeld}}, \bibinfo {author} {\bibfnamefont {M.~M.}\ \bibnamefont
  {Korshunov}}, \ and\ \bibinfo {author} {\bibfnamefont {I.~I.}\ \bibnamefont
  {Mazin}},\ }\href@noop {} {\bibfield  {journal} {\bibinfo  {journal} {Rep.
  Prog. Phys.}\ }\textbf {\bibinfo {volume} {74}},\ \bibinfo {pages}
  {124508} (\bibinfo {year} {2011})}\BibitemShut {NoStop}%
\bibitem [{\citenamefont {Kasahara}\ \emph {et~al.}(2012)\citenamefont
  {Kasahara}, \citenamefont {Hashimoto}, \citenamefont {Ikeda}, \citenamefont
  {Terashima}, \citenamefont {Matsuda},\ and\ \citenamefont
  {Shibauchi}}]{Kasahara2012}%
  \BibitemOpen
  \bibfield  {author} {\bibinfo {author} {\bibfnamefont {S.}~\bibnamefont
  {Kasahara}}, \bibinfo {author} {\bibfnamefont {K.}~\bibnamefont {Hashimoto}},
  \bibinfo {author} {\bibfnamefont {H.}~\bibnamefont {Ikeda}}, \bibinfo
  {author} {\bibfnamefont {T.}~\bibnamefont {Terashima}}, \bibinfo {author}
  {\bibfnamefont {Y.}~\bibnamefont {Matsuda}}, \ and\ \bibinfo {author}
  {\bibfnamefont {T.}~\bibnamefont {Shibauchi}},\ }\href@noop {} {\bibfield
  {journal} {\bibinfo  {journal} {Phys. Rev. B}\ }\textbf {\bibinfo {volume}
  {85}},\ \bibinfo {pages} {060503} (\bibinfo {year} {2012})}\BibitemShut
  {NoStop}%
\bibitem [{\citenamefont {Putzke}\ \emph {et~al.}(2012)\citenamefont {Putzke},
  \citenamefont {Coldea}, \citenamefont {Guillam\'on}, \citenamefont
  {Vignolles}, \citenamefont {McCollam}, \citenamefont {LeBoeuf}, \citenamefont
  {Watson}, \citenamefont {Mazin}, \citenamefont {Kasahara}, \citenamefont
  {Terashima}, \citenamefont {Shibauchi}, \citenamefont {Matsuda},\ and\
  \citenamefont {Carrington}}]{Putzke2012}%
  \BibitemOpen
  \bibfield  {author} {\bibinfo {author} {\bibfnamefont {C.}~\bibnamefont
  {Putzke}}, \bibinfo {author} {\bibfnamefont {A.~I.}\ \bibnamefont {Coldea}},
  \bibinfo {author} {\bibfnamefont {I.}~\bibnamefont {Guillam\'on}}, \bibinfo
  {author} {\bibfnamefont {D.}~\bibnamefont {Vignolles}}, \bibinfo {author}
  {\bibfnamefont {A.}~\bibnamefont {McCollam}}, \bibinfo {author}
  {\bibfnamefont {D.}~\bibnamefont {LeBoeuf}}, \bibinfo {author} {\bibfnamefont
  {M.~D.}\ \bibnamefont {Watson}}, \bibinfo {author} {\bibfnamefont {I.~I.}\
  \bibnamefont {Mazin}}, \bibinfo {author} {\bibfnamefont {S.}~\bibnamefont
  {Kasahara}}, \bibinfo {author} {\bibfnamefont {T.}~\bibnamefont {Terashima}},
  \bibinfo {author} {\bibfnamefont {T.}~\bibnamefont {Shibauchi}}, \bibinfo
  {author} {\bibfnamefont {Y.}~\bibnamefont {Matsuda}}, \ and\ \bibinfo
  {author} {\bibfnamefont {A.}~\bibnamefont {Carrington}},\ }\href@noop {}
  {\bibfield  {journal} {\bibinfo  {journal} {Phys. Rev. Lett.}\ }\textbf
  {\bibinfo {volume} {108}},\ \bibinfo {pages} {047002} (\bibinfo {year}
  {2012})}\BibitemShut {NoStop}%
\bibitem [{\citenamefont {Aichhorn}\ \emph {et~al.}(2011)\citenamefont
  {Aichhorn}, \citenamefont {Pourovskii},\ and\ \citenamefont
  {Georges}}]{Aichhorn2011}%
  \BibitemOpen
  \bibfield  {author} {\bibinfo {author} {\bibfnamefont {M.}~\bibnamefont
  {Aichhorn}}, \bibinfo {author} {\bibfnamefont {L.}~\bibnamefont
  {Pourovskii}}, \ and\ \bibinfo {author} {\bibfnamefont {A.}~\bibnamefont
  {Georges}},\ }\href@noop {} {\bibfield  {journal} {\bibinfo  {journal} {Phys.
  Rev. B}\ }\textbf {\bibinfo {volume} {84}},\ \bibinfo {pages} {054529}
  (\bibinfo {year} {2011})}\BibitemShut {NoStop}%
\bibitem [{\citenamefont {Blaha}\ \emph {et~al.}(2001)\citenamefont {Blaha},
  \citenamefont {Schwarz}, \citenamefont {Madsen}, \citenamefont {Kvasnicka},\
  and\ \citenamefont {Luitz}}]{Blaha2001}%
  \BibitemOpen
  \bibfield  {author} {\bibinfo {author} {\bibfnamefont {P.}~\bibnamefont
  {Blaha}}, \bibinfo {author} {\bibfnamefont {K.}~\bibnamefont {Schwarz}},
  \bibinfo {author} {\bibfnamefont {G.~K.~H.}\ \bibnamefont {Madsen}}, \bibinfo
  {author} {\bibfnamefont {D.}~\bibnamefont {Kvasnicka}}, \ and\ \bibinfo
  {author} {\bibfnamefont {J.}~\bibnamefont {Luitz}},\ }\href@noop {} {\emph
  {\bibinfo {title} {WIEN2k, An Augmented Plane Wave + Local Orbitals Program
  for Calculating Crystal Properties}}}\ (\bibinfo  {publisher} {Techn.
  Universit\"at Wien, Austria},\ \bibinfo {year} {2001})\BibitemShut {NoStop}%
\bibitem [{\citenamefont {Werner}\ \emph {et~al.}(2006)\citenamefont {Werner},
  \citenamefont {Comanac}, \citenamefont {de' Medici}, \citenamefont {Troyer},\
  and\ \citenamefont {Millis}}]{Werner2006}%
  \BibitemOpen
  \bibfield  {author} {\bibinfo {author} {\bibfnamefont {P.}~\bibnamefont
  {Werner}}, \bibinfo {author} {\bibfnamefont {A.}~\bibnamefont {Comanac}},
  \bibinfo {author} {\bibfnamefont {L.}~\bibnamefont {de' Medici}}, \bibinfo
  {author} {\bibfnamefont {M.}~\bibnamefont {Troyer}}, \ and\ \bibinfo {author}
  {\bibfnamefont {A.~J.}\ \bibnamefont {Millis}},\ }\href@noop {} {\bibfield
  {journal} {\bibinfo  {journal} {Phys. Rev. Lett.}\ }\textbf {\bibinfo
  {volume} {97}},\ \bibinfo {pages} {076405} (\bibinfo {year}
  {2006})}\BibitemShut {NoStop}%
\bibitem [{\citenamefont {Bauer}\ \emph {et~al.}(2011)\citenamefont {Bauer},
  \citenamefont {Carr}, \citenamefont {Evertz}, \citenamefont {Feiguin},
  \citenamefont {Freire}, \citenamefont {Fuchs}, \citenamefont {Gamper},
  \citenamefont {Gukelberger}, \citenamefont {Gull}, \citenamefont {Guertler},
  \citenamefont {Hehn}, \citenamefont {Igarashi}, \citenamefont {Isakov},
  \citenamefont {Koop}, \citenamefont {Ma}, \citenamefont {Mates},
  \citenamefont {Matsuo}, \citenamefont {Parcollet}, \citenamefont
  {Paw{\l}owski}, \citenamefont {Picon}, \citenamefont {Pollet}, \citenamefont
  {Santos}, \citenamefont {Scarola}, \citenamefont {Schollw\"ock},
  \citenamefont {Silva}, \citenamefont {Surer}, \citenamefont {Todo},
  \citenamefont {Trebst}, \citenamefont {Troyer}, \citenamefont {Wall},
  \citenamefont {Werner},\ and\ \citenamefont {Wessel}}]{Bauer2011}%
  \BibitemOpen
  \bibfield  {author} {\bibinfo {author} {\bibfnamefont {B.}~\bibnamefont
  {Bauer}}, \bibinfo {author} {\bibfnamefont {L.~D.}\ \bibnamefont {Carr}},
  \bibinfo {author} {\bibfnamefont {H.~G.}\ \bibnamefont {Evertz}}, \bibinfo
  {author} {\bibfnamefont {A.}~\bibnamefont {Feiguin}}, \bibinfo {author}
  {\bibfnamefont {J.}~\bibnamefont {Freire}}, \bibinfo {author} {\bibfnamefont
  {S.}~\bibnamefont {Fuchs}}, \bibinfo {author} {\bibfnamefont
  {L.}~\bibnamefont {Gamper}}, \bibinfo {author} {\bibfnamefont
  {J.}~\bibnamefont {Gukelberger}}, \bibinfo {author} {\bibfnamefont
  {E.}~\bibnamefont {Gull}}, \bibinfo {author} {\bibfnamefont {S.}~\bibnamefont
  {Guertler}}, \bibinfo {author} {\bibfnamefont {A.}~\bibnamefont {Hehn}},
  \bibinfo {author} {\bibfnamefont {R.}~\bibnamefont {Igarashi}}, \bibinfo
  {author} {\bibfnamefont {S.~V.}\ \bibnamefont {Isakov}}, \bibinfo {author}
  {\bibfnamefont {D.}~\bibnamefont {Koop}}, \bibinfo {author} {\bibfnamefont
  {P.~N.}\ \bibnamefont {Ma}}, \bibinfo {author} {\bibfnamefont
  {P.}~\bibnamefont {Mates}}, \bibinfo {author} {\bibfnamefont
  {H.}~\bibnamefont {Matsuo}}, \bibinfo {author} {\bibfnamefont
  {O.}~\bibnamefont {Parcollet}}, \bibinfo {author} {\bibfnamefont
  {G.}~\bibnamefont {Paw{\l}owski}}, \bibinfo {author} {\bibfnamefont {J.~D.}\
  \bibnamefont {Picon}}, \bibinfo {author} {\bibfnamefont {L.}~\bibnamefont
  {Pollet}}, \bibinfo {author} {\bibfnamefont {E.}~\bibnamefont {Santos}},
  \bibinfo {author} {\bibfnamefont {V.~W.}\ \bibnamefont {Scarola}}, \bibinfo
  {author} {\bibfnamefont {U.}~\bibnamefont {Schollw\"ock}}, \bibinfo {author}
  {\bibfnamefont {C.}~\bibnamefont {Silva}}, \bibinfo {author} {\bibfnamefont
  {B.}~\bibnamefont {Surer}}, \bibinfo {author} {\bibfnamefont
  {S.}~\bibnamefont {Todo}}, \bibinfo {author} {\bibfnamefont {S.}~\bibnamefont
  {Trebst}}, \bibinfo {author} {\bibfnamefont {M.}~\bibnamefont {Troyer}},
  \bibinfo {author} {\bibfnamefont {M.~L.}\ \bibnamefont {Wall}}, \bibinfo
  {author} {\bibfnamefont {P.}~\bibnamefont {Werner}}, \ and\ \bibinfo {author}
  {\bibfnamefont {S.}~\bibnamefont {Wessel}},\ }\href@noop {} {\bibfield
  {journal} {\bibinfo  {journal} {J. Stat. Mech.}\ ,\ \bibinfo {pages}
  {P05001}} (\bibinfo {year} {2011})}\BibitemShut {NoStop}%
\bibitem [{\citenamefont {Gull}\ \emph {et~al.}(2011)\citenamefont {Gull},
  \citenamefont {Werner}, \citenamefont {Fuchs}, \citenamefont {Surer},
  \citenamefont {Pruschke},\ and\ \citenamefont {Troyer}}]{Gull2011}%
  \BibitemOpen
  \bibfield  {author} {\bibinfo {author} {\bibfnamefont {E.}~\bibnamefont
  {Gull}}, \bibinfo {author} {\bibfnamefont {P.}~\bibnamefont {Werner}},
  \bibinfo {author} {\bibfnamefont {S.}~\bibnamefont {Fuchs}}, \bibinfo
  {author} {\bibfnamefont {B.}~\bibnamefont {Surer}}, \bibinfo {author}
  {\bibfnamefont {T.}~\bibnamefont {Pruschke}}, \ and\ \bibinfo {author}
  {\bibfnamefont {M.}~\bibnamefont {Troyer}},\ }\href@noop {} {\bibfield
  {journal} {\bibinfo  {journal} {Comp. Phys. Commun.}\ }\textbf {\bibinfo
  {volume} {182}},\ \bibinfo {pages} {1078 } (\bibinfo {year}
  {2011})}\BibitemShut {NoStop}%
\bibitem [{\citenamefont {Anisimov}\ \emph {et~al.}(1997)\citenamefont
  {Anisimov}, \citenamefont {Aryasetiawan},\ and\ \citenamefont
  {Lichtenstein}}]{Anisimov1997}%
  \BibitemOpen
  \bibfield  {author} {\bibinfo {author} {\bibfnamefont {V.~I.}\ \bibnamefont
  {Anisimov}}, \bibinfo {author} {\bibfnamefont {F.}~\bibnamefont
  {Aryasetiawan}}, \ and\ \bibinfo {author} {\bibfnamefont {A.~I.}\
  \bibnamefont {Lichtenstein}},\ }\href@noop {} {\bibfield  {journal} {\bibinfo
   {journal} {J. Phys.: Condens. Matter}\ }\textbf {\bibinfo {volume} {9}},\
  \bibinfo {pages} {767} (\bibinfo {year} {1997})}\BibitemShut {NoStop}%
\bibitem [{\citenamefont {Czy\ifmmode~\dot{z}\else \.{z}\fi{}yk}\ and\
  \citenamefont {Sawatzky}(1994)}]{Czyzyk1994}%
  \BibitemOpen
  \bibfield  {author} {\bibinfo {author} {\bibfnamefont {M.~T.}\ \bibnamefont
  {Czy\ifmmode~\dot{z}\else \.{z}\fi{}yk}}\ and\ \bibinfo {author}
  {\bibfnamefont {G.~A.}\ \bibnamefont {Sawatzky}},\ }\href@noop {} {\bibfield
  {journal} {\bibinfo  {journal} {Phys. Rev. B}\ }\textbf {\bibinfo {volume}
  {49}},\ \bibinfo {pages} {14211} (\bibinfo {year} {1994})}\BibitemShut
  {NoStop}%
\bibitem [{\citenamefont {de' Medici}(2011)}]{Medici2011}%
  \BibitemOpen
  \bibfield  {author} {\bibinfo {author} {\bibfnamefont {L.}~\bibnamefont {de'
  Medici}},\ }\href@noop {} {\bibfield  {journal} {\bibinfo  {journal} {Phys.
  Rev. B}\ }\textbf {\bibinfo {volume} {83}},\ \bibinfo {pages} {205112}
  (\bibinfo {year} {2011})}\BibitemShut {NoStop}%
\bibitem [{\citenamefont {Borisenko}\ \emph {et~al.}(2010)\citenamefont
  {Borisenko}, \citenamefont {Zabolotnyy}, \citenamefont {Evtushinsky},
  \citenamefont {Kim}, \citenamefont {Morozov}, \citenamefont {Yaresko},
  \citenamefont {Kordyuk}, \citenamefont {Behr}, \citenamefont {Vasiliev},
  \citenamefont {Follath},\ and\ \citenamefont {B\"uchner}}]{Borisenko2010}%
  \BibitemOpen
  \bibfield  {author} {\bibinfo {author} {\bibfnamefont {S.~V.}\ \bibnamefont
  {Borisenko}}, \bibinfo {author} {\bibfnamefont {V.~B.}\ \bibnamefont
  {Zabolotnyy}}, \bibinfo {author} {\bibfnamefont {D.~V.}\ \bibnamefont
  {Evtushinsky}}, \bibinfo {author} {\bibfnamefont {T.~K.}\ \bibnamefont
  {Kim}}, \bibinfo {author} {\bibfnamefont {I.~V.}\ \bibnamefont {Morozov}},
  \bibinfo {author} {\bibfnamefont {A.~N.}\ \bibnamefont {Yaresko}}, \bibinfo
  {author} {\bibfnamefont {A.~A.}\ \bibnamefont {Kordyuk}}, \bibinfo {author}
  {\bibfnamefont {G.}~\bibnamefont {Behr}}, \bibinfo {author} {\bibfnamefont
  {A.}~\bibnamefont {Vasiliev}}, \bibinfo {author} {\bibfnamefont
  {R.}~\bibnamefont {Follath}}, \ and\ \bibinfo {author} {\bibfnamefont
  {B.}~\bibnamefont {B\"uchner}},\ }\href@noop {} {\bibfield  {journal}
  {\bibinfo  {journal} {Phys. Rev. Lett.}\ }\textbf {\bibinfo {volume} {105}},\
  \bibinfo {pages} {067002} (\bibinfo {year} {2010})}\BibitemShut {NoStop}%
\bibitem{Note1}%
  \bibinfo {note} {C. Putzke, private communication.}%

\end{thebibliography}


%

\end{document}